# A material decomposition method for dual-energy CT via dual interactive Wasserstein generative adversarial networks

Zaifeng Shi[a,c,*], Huilong Li[a], Qingjie Cao[b], Zhongqi Wang[a], Ming Cheng[a]

[a] *School of Microelectronics, Tianjin University, Tianjin 300072, China*

[b] *School of Mathematical Sciences, Tianjin Normal University, Tianjin 300072, China*

[c] *Tianjin Key Laboratory of Imaging and Sensing Microelectronic Technology, Tianjin 300072, China*

**Purpose:** Dual-energy computed tomography (DECT) is highly promising for material characterization and identification, whereas reconstructed material-specific images are affected by magnified noise and beam-hardening artifacts. Although various DECT material decomposition methods have been proposed to solve this problem, the quality of the decomposed images is still unsatisfactory, particularly in the image edges. In this study, a data-driven approach using a dual interactive Wasserstein generative adversarial network (DIWGAN) is developed to improve DECT decomposition accuracy and perform edge-preserving images.

**Methods:** In the proposed DIWGAN, two interactive generators are used to synthesize the decomposed images of two basis materials by modeling the spatial and spectral correlations from input DECT reconstructed images, and the corresponding discriminators are employed to distinguish the difference between the generated images and labels. The DECT images reconstructed from high- and low-energy bins are sent to two generators separately, and each generator synthesizes one material-specific image, thereby ensuring the specificity of the network modeling. In addition, the information from different energy bins is exploited through the feature sharing of two generators. During decomposition model training, a hybrid loss function including $L_1$ loss, edge loss, and adversarial loss are incorporated to preserve the texture and edges in the generated images. Additionally, a selector is employed to define the generator that should be trained in each iteration, which can ensure the modeling ability of two different generators and improve the material decomposition accuracy. The performance of the proposed method is evaluated using a digital phantom, the XCAT phantom, and real data from a mouse.

---
[*] Corresponding author.
Email address: shizaifeng@tju.edu.cn

**Results:** On the digital phantom, the regions of bone and soft tissue are strictly and accurately separated using the trained decomposition model. The material densities in different bone and soft-tissue regions are near the ground truth, and the error of material densities is lower than 3 mg/ml. Compared with Butterfly-Net, the root-mean-square error (RMSE) of soft-tissue images generated by the DIWGAN decreased by 0.01 g/mL, whereas the peak-signal-to-noise ratio (PSNR) and structural similarity (SSIM) of the soft-tissue images reached 31.43 dB and 0.9987, respectively. The mass densities of the decomposed materials are nearest to the ground truth when using the DIWGAN method. The noise standard deviation of the decomposition images reduced by 69%, 60%, 33%, and 21% compared with direct matrix inversion, iterative decomposition, fully convolutional network, and Butterfly-Net, respectively. Furthermore, the performance of the mouse data indicates the effectiveness of the proposed material decomposition method in real scanned data.

**Conclusions:** A DECT material decomposition method based on deep learning is proposed, and the relationship between reconstructed and material-specific images is mapped by training the DIWGAN model. Results from both the simulation phantoms and real data demonstrate the advantages of this method in suppressing noise and beam-hardening artifacts.

Keywords**:** Dual-energy CT, material decomposition, generative adversarial network, loss function

## 1. INTRODUCTION:

X-ray computed tomography (CT) can provide tomographic images of human tissues[1] and is typically used in clinical practice and medical diagnosis. Compared with conventional CT, dual-energy computed tomography (DECT) can acquire additional diagnostic information by utilizing two different spectra. Hence, DECT can decompose material-specific images, known as material decomposition, which has many advanced imaging applications, such as contrast agent quantification,[2] abdomen angiography detection[3], and kidney stone characterization.[4,5] Compton scattering and photoelectric absorption can be combined linearly to estimate the linear attenuation coefficient of each voxel in reconstructed images. Because the physical effect is related to the elemental atomic number, energy-specific attenuation can be represented by the combined attenuation of two basis materials.

Pioneered by Alvarez and Macovski[6] in 1976, various approaches for DECT material decomposition have been proposed, and they can be classified into three categories: direct reconstruction, raw-data-based, and image-based methods. Direct reconstruction methods incorporate the dual-energy CT transmission and material decomposition models to realize decomposition and reconstruction simultaneously.[7] However, their computational cost is high, and their decomposition accuracy is sensitive to the system configuration. Regarding raw-data-based methods, projections from detectors are decomposed into material-specific sinograms initially, and then the material images can be obtained by conventional CT reconstruction algorithms such as filtered back projection (FBP).[8,9] The decomposition accuracy of these methods is often affected significantly by the mismatch of the projection data in different energy bins. Therefore, the acquired projections must exhibit spatial geometric consistency, which is challenging in current DECT systems. Unlike decompositions based on raw data, image-based methods employ reconstructed images directly and obtain decomposed images according to the linear attenuation coefficient of each pixel.[10,11] The images acquired from commercial DECT scanners can be conveniently used compared with the other two methods. Nevertheless, noise magnification during decomposition is a persistent problem, and many approaches have been proposed for image-based material decomposition. For instance, Niu et al. proposed an iterative approach to suppress DECT image noise by considering the full variance–covariance matrix of material images.[12] To obtain a more uniform noise power spectrum for DECT, Harms et al. proposed a framework for penalized weighted least-squares optimization with similarity-based regularization[13]. To fully utilize prior knowledge, Xu et al.[14] developed the dictionary learning method to improve the quality of low-dose CT images. Subsequently, this approach was employed in DECT for image reconstruction and decomposition.[15,16] It was proven effective for exploiting learned or adapted prior information to suppress decomposition noise and artifacts.[17,18] However, owing to the noise and artifacts in the reconstructed images, it is difficult to map the relationship between the images under different energy bins and the basis material images in these conventional methods. Hence, more advanced methods must be developed to further improve the material decomposition accuracy.

Deep learning, with its effective feature extraction and modeling ability, has been used in recent years in various medical imaging applications, such as image segmentation, classification, and recognition. A perspective article from Wang indicated that various tomographic modalities are representable when using deep learning.[19] Subsequently, an increasing number of medical image analysis methods have been proposed, including noise reduction of low-dose CT images,[20] sparse-view CT reconstruction,[21,22] and organ segmentation.[23] In addition, some researchers have focused on applying deep learning to material decomposition. Badea et al. proposed the use of convolutional neural networks (CNNs) for spectral micro-CT material decomposition and achieved satisfactory results.[24] Additionally, Xu et al. adopted a fully convolutional network (FCN) to decompose DECT images[25]. To establish a tight connection between a decomposition model and a network, Zhang et al.[26] designed a butterfly network (Butterfly-Net) to fully optimize the performances of networks and achieved promising material decomposition. In 2014, Goodfellow et al. proposed a novel network architecture, known as the generative adversarial network (GAN),[27] which can preserve texture details more effectively and produce visually appealing images by adversarial training. For example, to suppress noise caused by dose reduction, Wolterink et al. used the GAN to generate normal-dose CT images from low-dose CT images.[28] Compared with traditional CNN methods, it avoided the blur effect on the resultant denoised images. However, GANs are typically affected by mode collapse and exhibit convergence problems; therefore, different loss functions for the discriminator have been proposed, such as least square,[29] f-divergence,[30] and the Wasserstein distance.[31,32] Among these GAN variants, the Wasserstein GAN is arguably the most popular; it not only solves the problem of unstable training, but also provides a reliable indicator of the training process. In addition, employing two generators[33] or discriminators[34] in the adversarial training process have been proven effective in solving the mode collapse problem. Furthermore, the loss function of the generator is essential in network learning, and different loss functions or their combinations have been used to accomplish specific learning tasks efficiently.[35] For example, Yang et al. coupled the mean squared error (MSE) loss of both image and frequency domains as well as a perceptual loss to provide superior reconstruction images for Magnetic Resonance Imaging.[36]

Inspired by the promising performance of adversarial training, an image-based material decomposition method using dual interactive Wasserstein generative adversarial network (DIWGAN) was proposed to solve the problem of noise magnification and to remove beam-hardening artifacts from DECT material decomposition. In contrast to the current CNN-based material decomposition methods, two interactive generators are used in the DIWGAN to synthesize the corresponding material images and exchange information from different energy bins. Two generators are trained with a hybrid loss to improve the decomposition performance of image texture and edges. Additionally, a selector is employed for training the generators alternatively, which can prevent the two generators from being trapped in a local optimum.

The remainder of this paper is organized as follows: in Section 2, the basics of DECT material decomposition and detailed information regarding the proposed network are provided. The corresponding experiments and results are presented in Section 3. The discussion and conclusion are presented in Sections 4 and 5, respectively.

## 2. MATERIALS AND METHODS:

### 2.A. Image-based material decomposition for DECT

For image-based material decomposition, the input data comprises a pair of reconstructed images from two energy bins. According to image-based decomposition theory,[37] the linear combination of pixel values in the basis material images can represent the linear attenuation coefficient of each pixel in the input images. Suppose the total number of pixels in a reconstructed image is $n$, and the two basis material images can be obtained from the reconstructed DECT images by a direct matrix inversion, which is written as follows:

$$\mathbf{x} = \mathbf{A}^{-1}\boldsymbol{\mu}, \qquad (1)$$

where $\boldsymbol{\mu}$ and $\mathbf{x}$ are *2n* vectors comprising reconstructed and decomposed images, respectively, and $\mathbf{A}^{-1}$ represents the decomposition matrix. However, noise always exists in reconstructed images; therefore, it is difficult to accurately obtain the linear attenuation coefficient of different materials in $\mathbf{A}^{-1}$. Consequently, noise will be increased significantly while obtaining material-specific images, and this can be treated as an ill-posed problem. Inspired by the powerful mapping ability of deep learning, we propose using a mapping function $F$ to realize material decomposition for DECT. Therefore, the relationship between the reconstructed and decomposed images can be expressed as follows:

$$\mathbf{x} = F(\boldsymbol{\mu}) \tag{2}$$

where $F$ is approximated using a trained network model (i.e., the DIWGAN) in our method.

## 2.B. Material decomposition based on DIWGAN

For DECT material decomposition, the data distribution of the two basis materials must be generated from the reconstructed images under two different energy bins. The two basis materials will always have different data distributions, whereas the reconstructed images from high- and low-energy bins are significantly related. Therefore, two generators were used in the proposed approach for different targets, and the correlations of the reconstructed images were fully utilized through the feature exchange between them. Inspired by the butterfly network,[26] which realized feature sharing by one crossover architecture based on residual learning, we employed three crossover architectures between two generators. The network structure of the generators was based on U-net[38]; therefore, the tissue features from three different levels can be shared during the decomposition. An overview of the proposed network for DECT material decomposition is shown in Fig. 1.

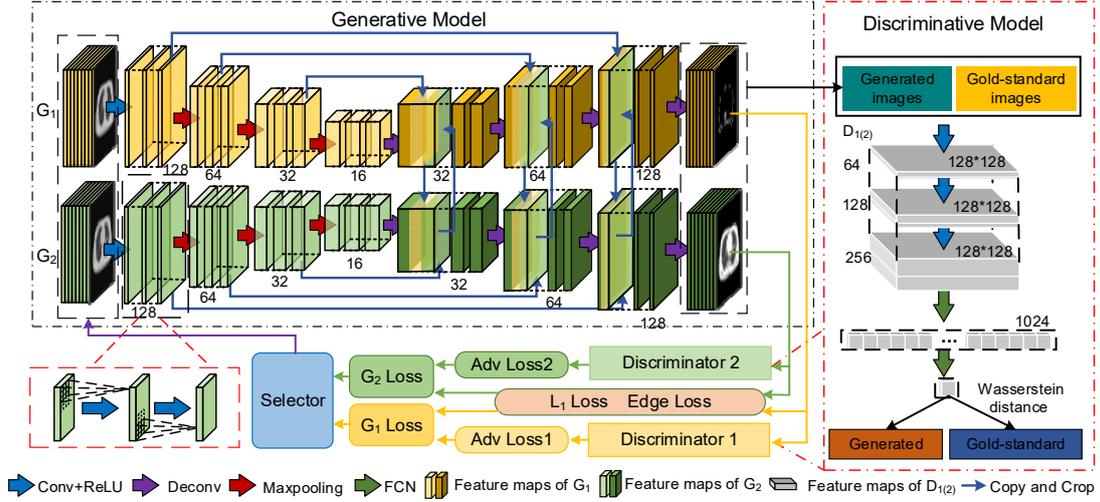

Fig. 1. Overview of proposed DECT material decomposition method.

The images from high - and low-energy bins were sent to two generators ($G_1$, $G_2$) separately, and each generator synthesized one material-specific image, which can ensure the specificity of the network modeling. In addition, $G_1$ and $G_2$ interacted with each other to share information from different input data such that both the spatial and spectral correlations were learned during the training. To ensure that these two generators yielded accurate material-specific images, the corresponding discriminators ($D_1$, $D_2$), restricted by the discriminator loss, were adopted in the proposed network. The data distributions of the ground truth ($P_r$) and the generated image ($P_g$) were compared using the Wasserstein distance instead of the JS divergence used in the original GAN[32]. The min–max problem between $G_{1(2)}$ and $D_{1(2)}$ can be described as follows:

$$
\begin{aligned}
\min_{G_{1(2)}} \max_{D_{1(2)}} L_{WGAN}(D_{1(2)}, G_{1(2)}) = &-E_{Y \sim P_r}[D_{1(2)}(Y)] + E_{X \sim P_p}[D_{1(2)}(G_{1(2)}(X))] \\
&+ \lambda E_{\hat{Y} \sim P_g}[(\| \nabla_{\hat{Y}} D_{1(2)}(\hat{Y}) \|_2 - 1)^2]
\end{aligned}
\tag{3}
$$

where the first two terms denote the Wasserstein distance estimation. $X$ represents the input image with data distribution $P_x$, and $Y$ represents the ground truth. The last term is the regularization term, and $\lambda$ represents the penalty coefficient. $\hat{Y}$ is generated by uniformly sampling the corresponding synthetic and real samples along a straight line. Through adversarial training, the generators synthesized decomposed material images that were the same as the real material-specific images. In the proposed method, the differences between the target and the generated images were lessened by a hybrid loss that included $L_1$, edge, and adversarial losses. Owing to the differences between the two basis materials, the hybrid loss values of $G_1$ and $G_2$ differed during the network training. If two generators were trained together, then the $G_1$ and $G_2$ losses would have the same weight in the backpropagation process; therefore, the network performance would be determined by the loss with the greater value, which will result in convergence problems. Inspired by a previous study,[39] a selector (S) was employed to solve this problem. Using the values of the hybrid loss as input, S can decide the generator to be trained in the next iteration. Hence, the $G_1$ and $G_2$ losses can obtain the adaptive weights during the training and converge them to the minimum rather than a compromise, thereby guaranteeing the mapping ability of the generators. A specific description of the proposed network structure is provided in the following.

### 2.B.1. Generative model

As shown in Fig. 1, the two generators $G_1$ and $G_2$ were constructed with the same architecture because both aim to synthesize a material-specific image from a reconstructed image. This encoder–decoder architecture comprises of two parts: a contracting path and an expansive path. The two generators focused on extracting the features from the corresponding energy bins independently in the contracting path and realize information exchange and material decomposition in the expansive path. The numbers of fitters in the convolutional layer were 64, 128, 256, and 512, which were used to extract the features from different levels. The kernel size was $3 \times 3$, with a stride of one unit. Max pooling layers were adopted to reduce the size of maps, thereby allowing the network to obtain coarse features and prevent overfitting. The kernel size and the stride of the pooling layer were $2 \times 2$ and 2, respectively. In the expansive path, deconvolution with a $3 \times 3$ kernel and stride 2 was performed to preserve the spatial information and ensure that the sizes of the input and output images matched. Specifically, the architecture of each generator was based on U-net, and the features were classified into three levels by the pooling layers in the contracting path. Copy and cropping were performed to add these features from one generator to the other; therefore, in the expansive path, the features from two energy bins can be utilized to generate material-specific images. It is noteworthy that the information exchange between images from the high- and low-energy bins is necessary for DECT material decomposition. Generally, the reconstructed images from low-energy bins contain many detailed texture information of the tissues, whereas some small features are invisible in the high-energy images. In addition, the noise distribution differs, and by cropping the feature maps learned from the two energy bins, the network can fully utilize the information from different energies, thereby significantly improving the robustness and accuracy of the decomposition model. Near the output layer, a 3 × 3 kernel with one channel was used to match the dimensions of the material-specific images. All the convolutional layers were proceeded by an activation function, known as the rectified linear unit (ReLU),[40] which was used to improve the fitting ability of the network.

### 2.B.2. Discriminative model

The architecture of the proposed discriminator shared by $D_1$ and $D_2$ is shown in Fig. 1. The discriminator uses either the generated material-specific images or the gold-standard decomposed images as input and judges whether the input is ideal material-specific images. To capture both the low- and high-level features of the input, six convolutional layers were used for feature extraction, and every two contained the same numbers of filters, which were 64, 128, and 256. All the kernel sizes were 3 × 3 with a stride of 1 pixel, and ReLU was added after each convolutional layer. Near the output, two fully connected layers were adopted in the architecture for feature integration and expression, and the numbers of units were 1024 and 1, respectively. In addition, the discriminator was trained by estimating the Wasserstein distance of the generated images and ground truth from the output such that the corresponding generator would yield accurate decomposed images. $D_1$ and $D_2$ were employed for two different material images, and the probability of the input images being considered as the real decomposed images was used as part of the loss functions of $G_1$ and $G_2$, respectively.

### *2.B.3 Selector*

The selector used the values of loss functions from $G_1$ and $G_2$ as input and determined the generator that should be trained in the next iteration. Let $L_{c1}$ and $L_{c2}$ represent the values that $G_1$ loss ($L_{g1}$) and $G_2$ loss ($L_{g2}$) will approach during the training, respectively. To select a suitable value for $L_{c1}$ and $L_{c2}$, two generators were first trained synchronously and after one of the generator losses had converged, the initial values of $L_{c1}$ and $L_{c2}$ were selected based on the average values of $G_1$ and $G_2$ losses from three consequent epochs, respectively. Subsequently, $L_{c1}$ and $L_{c2}$ were adjusted according to the convergence of $G_1$ and $G_2$ losses via multiple experiments. The distance ($d_k$) between the losses of the generators and the corresponding objective values can be expressed as follows:

$$d_k = \left| L_{gk} - L_{ck} \right| (k = 1, 2),$$ (4)

where the subscript $k$ represents the two generators $G_1$ and $G_2$. By comparing the values of $d_1$ and $d_2$, the performances of the two generators were evaluated during each iteration. In the next iteration, the generator with a larger distance was trained. This training strategy ensures that each generator obtains the adaptive training times and improves the training efficiency of the DIWGAN.

### *2.B.4 Hybrid Loss function*

To improve the material decomposition performance of the DIWGAN, various factors should be considered during network training, such as the value in each voxel, image edges, and texture details. Therefore, a hybrid loss including L₁, edge, and adversarial losses was proposed to supervise DIWGAN learning.

The L₁ loss is a mean-based measure, which is also known as the mean absolute error. It ensures that each voxel of the generated material-specific images has an accurate correspondence to the ground truth. Compared with the typically used L₂ loss, the L₁ loss does not excessively punish large differences or tolerate small errors; additionally, it can maintain the same fine characteristics.[41] The L₁ loss can be expressed as follows:

$$L_1 = \frac{1}{hwd} \left| G(X) - Y \right|, \qquad (5)$$

where $h$, $w$, and $d$ represent the height, width, and depth of the input reconstructed images, respectively; G(X) represents the synthetic images from the generators; $Y$ represents the ground truth. In addition, because the basis material images always have common or complementary boundaries, the edge information cannot always be well preserved during the decomposition. We propose using the edge loss to improve the performance, which can be expressed as follows:

$$L_{edge} = \left\| \left| \nabla Y_x \right| - \left| \nabla X_x \right| \right\|^2 + \left\| \left| \nabla Y_y \right| - \left| \nabla X_y \right| \right\|^2 + \left\| \left| \nabla Y_z \right| - \left| \nabla X_z \right| \right\|^2, \qquad (6)$$

where the subscripts $x$, $y$, and $z$ represent the directions of gradient descent of the input images X and ground truth Y. This loss attempts to minimize the magnitudes of the gradients between the generated images and the ground truth. Hence, the edges that always exhibit a strong gradient can be well preserved during the minimizing process. Moreover, as an essential aspect in adversarial training, adversarial loss enables the generator to synthesize decomposed material images that are similar to the real material-specific images. Unlike the $L_1$ and edge losses, the adversarial loss can eliminate the requirement of modeling explicit pixel-wise objective functions. Instead, a rich similarity metric is learned to distinguish between real and fake data, which optimizes the concepts beyond the pixel level in images, resulting in more realistic results. The adversarial loss is defined as follows:

$$L_{adv} = \min_{G} \max_{D} L_{WGAN}(D, G) \qquad (7)$$

This min–max optimization framework enables generators to provide the same high-level features in decomposed images as in the ground truth. The texture information of fine tissue structures will be well preserved by minimizing the adversarial loss.

By combining all the losses above, the hybrid loss can be expressed as follows:

$$L_g = \lambda_1 L_1 + \lambda_2 L_{edge} + \lambda_3 L_{adv}, \qquad (8)$$

where $\lambda_1$, $\lambda_2$, and $\lambda_3$ denote the weights of the different losses; these parameters were set according to the scales of different loss terms and adjusted based on the training results. Through the linear combination of these terms, the beneficial performance afforded by various losses can be achieved simultaneously, and the generator can yield high-quality material-specific images by minimizing the hybrid loss.

## 2.C. Experimental datasets and setup

### 2.C.1 Data acquisition and evaluation

To test and validate the proposed network in DECT material decomposition, a number of datasets were used to train the decomposition model and evaluate its performance. In this experiment, real human phantoms from 4D XCAT from Duke University[42] were randomly drawn as training and testing data. To improve the robustness of the proposed network, 10 patients of different genders, ages, heights, and weights were recruited for the experiment, and the human phantoms used were non-enhanced. The Edge-on X-ray detector model proposed previously[43] was used to provide data from two different energy bins. It operated as a dual-layer detector, and the absorption, scattering, and random noise of different levels were simulated in the adsorption process of photons. A GE_Maxiray_125 tube operated at 140 kVp was applied to produce X-ray photons, and Sidden's ray-driven algorithm[44] was employed to simulate the fan-beam geometry. The distance between the source-to-rotation center and detector-to-rotation center was set to 59.5 cm. The image region measured 33 cm × 33 cm, and the number of views over 360° was 360. Because the entire DECT scanning process was based on simulation, the acquired system matrix did not require further calibrations in the experiment. In addition, the datasets of mice acquired by spectral micro-CT from MARS[45] were used to evaluate the potential of the proposed method in real scanned data. The reconstructed images of the mouse head measuring 256 cm × 256 cm from 36–52 and 52–80 keV energy bins were selected, and the pixel size was 110 μm × 110 μm.

In the experiment, images of the human phantoms were reconstructed using the FBP algorithm. To satisfy the clinical geometry, the energy threshold of the Edge-on detector was set to 80 keV, and because the photons under 20 keV were few, 0 to 20 keV was considered as the auxiliary energy bin in the experiment. Hence, the two energy bins were allocated as 20–80 and 80–140 keV. To obtain the training label, two ideal monochromatic images under 50 and 110 keV of one slice were derived from XCAT, and an algorithm known as direct matrix inversion was used to obtain the material-specific images by decomposing the input ideal images. Compared with using polychromatic images to obtain the label, the linear attenuation coefficient of different tissues in the monochromatic images is more precise, and the monochromatic images can effectively reduce beam-hardening artifacts. In total, 640 pairs of images with 256 × 256 pixels under the two energy bins and the corresponding labels were derived from different slices of eight patients. In addition, 160 pairs of slices from the remaining two patients were obtained to test the decomposition performance. These slices were extracted from both the lungs and heads, and bone and soft tissue were the two basis materials. To further improve the size of the training datasets and prompt the network to learn more detailed information, we partitioned the input images and labels into overlapping 128 × 128 image patches, and the sliding interval was 16 pixels. Finally, we obtained 518400 pairs of image patches for training.

To evaluate the decomposition performance and effectiveness of the proposed network, a digital phantom filled with bone and soft tissue, phantom from XCAT, and real mouse dataset were used in the experiment. In addition, four typical approaches, including direct matrix inversion, iterative decomposition,[12] FCN method,[24] and Butterfly-Net[26] were used for comparison. The PSNR, SSIM, and RMSE were calculated for a quantitative evaluation. To further evaluate the statistical performance of different decomposition approaches, the mean and standard deviation (SD) in different regions of interests (ROIs) were calculated and compared. The RMSE was calculated using the mass density, which is more straightforward for evaluating the degree of quantitative accuracy; other indicators in the experiment were obtained based on the values of volume fractions.

### 2.C.2 Network training and convergence

In the experiments, all the generators and discriminators were optimized using the Adam algorithm,[46] and the learning rate was set to $1 \times 10^{-4}$ with two hyper-parameters, $\beta_1 = 0.5$ and $\beta_2 = 0.9$. Following a previous suggestion[32], the parameter $\lambda$ was set to 10, which was used to balance the Wasserstein distance and the regularization term. During the training, the mini-batch size was set to 32. According to our experience, $\lambda_1 = 1$, $\lambda_2 = 0.5$, and $\lambda_3 = 0.01$ were selected to balance the weights of different losses; and $L_{c1}$ and $L_{c2}$ were set to 10 and 8, respectively to balance the training of the two generators. The network was implemented on an Nvidia RTX 2080 GPU based on the Tensorflow framework.[47] To visualize the training process of the network, the training and validation curves of $G_1$ and $G_2$ losses are plotted in Figs. 2 (a) and (b), respectively. First, the training curves of both generators decreased rapidly, indicating that the learning rate was suitable, and the gradient descent process was performed. After the network iterated for 800 epochs, both curves became smooth and converged to a minimum. The trend of the validation curves was the same as that of the training curves, and their values were similar, indicating the good fitting ability of the network model. The total training time of the proposed network model was approximately 45 h.

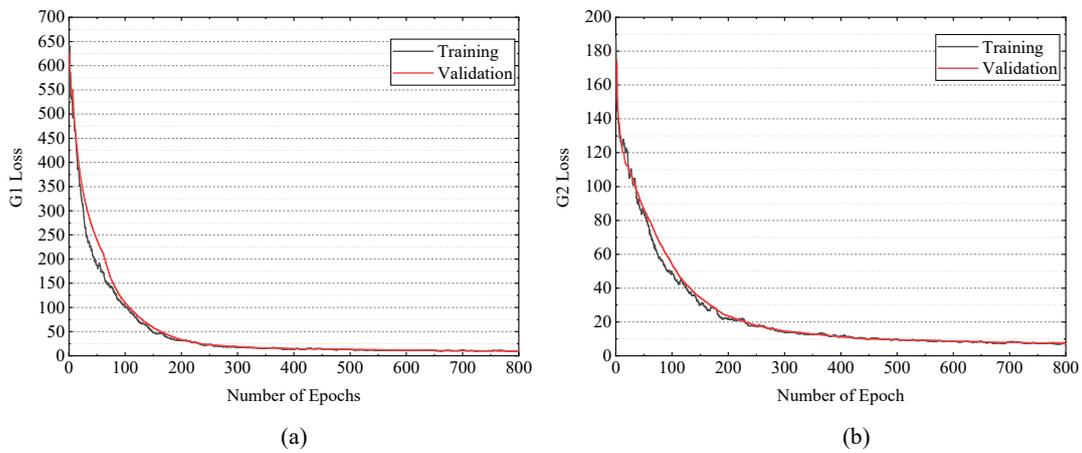

Fig. 2. Training and validation curves of $G_1$ and $G_2$ losses.

# 3. Results:

## 3.A. Digital phantom study

A digital phantom with different material densities was constructed to evaluate the decomposition accuracy and robustness of the trained decomposition model. The reconstructed images from both the high- and low-energy bins were obtained similarly to generate the training data. Figure 3(a) and (b) show the reconstructed images from the 20–80 and 80–140 keV energy bins, respectively. The bone and soft-tissue regions are marked in Fig. 3(a), and the specific densities are shown in Table 1.

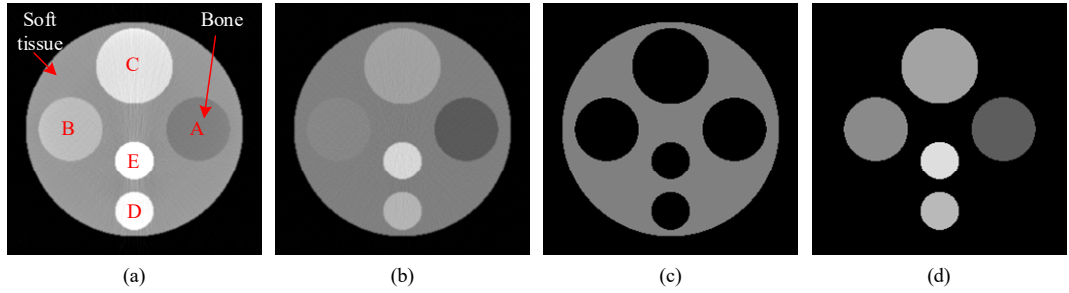

Fig. 3 Results of digital phantom evaluation. (a) and (b) show reconstructed images from 20–80 and 80–140 keV energy bins. (c) and (d) show soft tissue and bone images generated by our proposed network. The display window size is [0 0.5].

**Table 1**
A list of material densities on different regions generated by DIWGAN.

| Region | Ground Truth(g/ml) | DIWGAN(g/ml) |
|---|---|---|
| Soft tissue | 1.500 | 1.497 |
| Bone A | 1.000 | 1.002 |
| Bone B | 1.500 | 1.500 |
| Bone C | 1.800 | 1.799 |
| Bone D | 2.000 | 2.003 |
| Bone F | 2.400 | 2.402 |

The decomposition results of the bone and soft tissue from the DIWGAN are shown in Figs. 3 (c) and (d), respectively. The regions of bone with different densities and soft tissue were strictly and accurately separated, and the noise in the input images was invisible in the generated material images. In addition, to quantitatively evaluate the decomposition accuracy of the DIWGAN, the bone and soft-tissue densities in the generated images were tested. As shown in Table 1, the material densities in different bone and soft-tissue regions were near the ground truth, and the error of the material densities was lower than 3 mg/mL, demonstrating the quantitative decomposition accuracy of our trained network model.

### 3.B. XCAT phantom study

To test the effectiveness of the proposed material decomposition model, the slices from the test datasets were selected for qualitative and quantitative comparisons. The decomposition results from different methods are shown in Fig. 4. The first column is the ground truth of the decomposed images for bone and soft tissue. Figures 4 (b1)–(b4) show the material images generated by direct matrix inversion; as shown, noise was apparent because this method only linearly combined the images from high- and low-energy bins, resulting in magnified noise levels. As shown in Figs. 4 (c1)–(c4), the iterative decomposition suppressed the noise and artifacts, but it delivered a smooth image because of the smoothness regularization term in the decomposition function. In addition, beam-hardening artifacts remained visible, particularly in the soft-tissue images of the head. This may be because these conventional material decomposition algorithms are only responsible for decomposition and do not contain any beam-hardening corrections. We discovered that the FCN method removed most artifacts, and that noise was invisible in the material images, as shown in Figs. 4 (d1)–(d4). Regarding the Butterfly-Net shown in Figs. 4 (e1)–(e2), the noise and beam-hardening artifacts were also invisible. However, the performance of edge preservation was unsatisfactory compared with that of the ground truth because only the MSE was used as the loss function in the FCN and Butterfly-Net, thereby yielding burring edge images during network training. Figures 4 (f1)–(f4) show that both the bone and soft-tissue images generated by our proposed method were visually similar to the standard images. Compared with the comparisons above, our decomposition model yielded clear material image edges with the least amount of noise and artifacts. To further clarify this point, the corresponding ROIs (A and B) of the soft tissue marked with red rectangles in Fig. 4 (a2, a4) of different material decomposition methods were extracted and magnified, as shown in Fig. 5. The first column shows the targets; it is clear that the soft-tissue image quality degraded by the direct matrix inversion, as shown in Figs. 5 (b1) and (b2). Using the iterative decomposition method, beam-hardening artifacts were observed in the soft-tissue images of the head, as shown in Fig. 5 (c2). On the contrary, the results from Figs. 5 (d1, d2), (e1, e2), and (f1, f2) further confirmed that the learning-based methods are promising in avoiding noise magnification. Moreover, comparing Fig. 5 (f1) with Figs. 5(d1) and (e1), the proposed DIWGAN can provide clear image edges and structures, which will be beneficial in distinguishing anatomical structures.

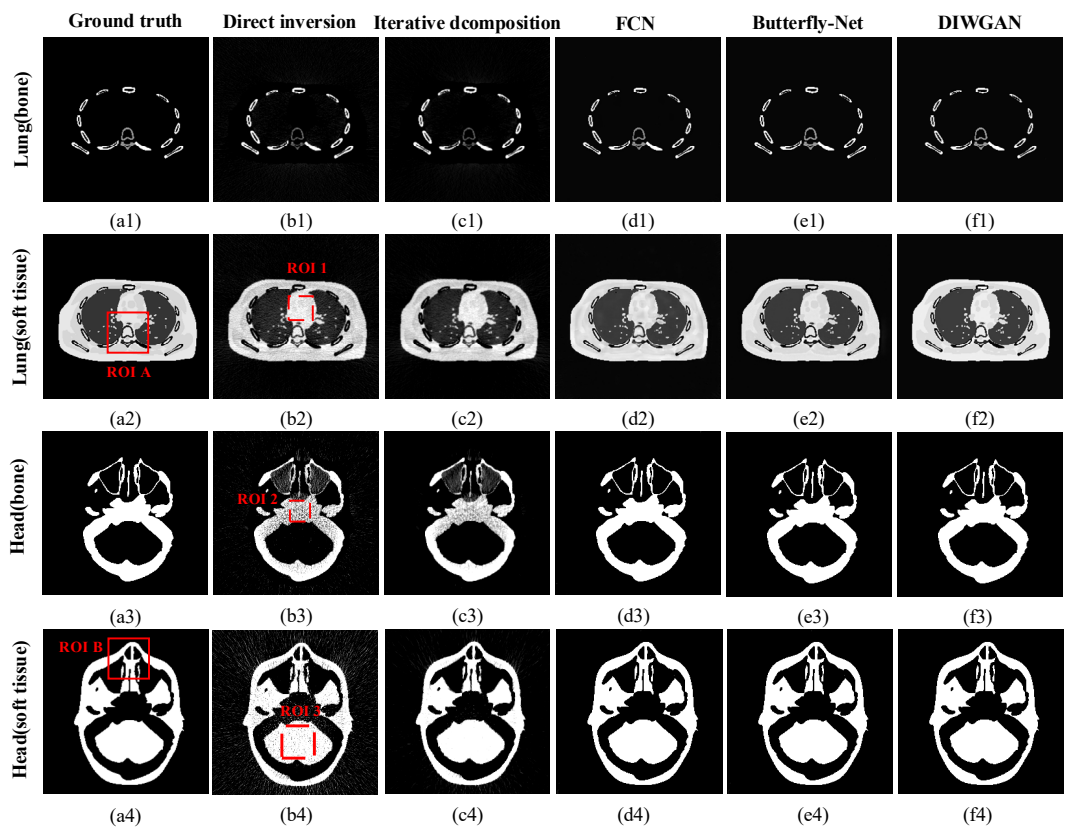

Fig. 4. Decomposed bone and soft-tissue images of lung and head by different methods. The display window sizes of lung and head are [0, 1.5] and [0.1, 1.1], respectively.

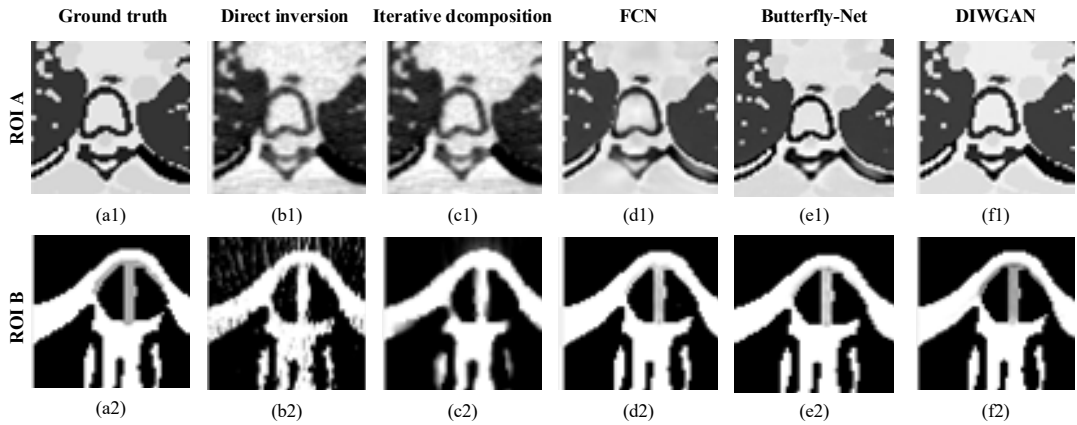

Fig. 5. Magnified ROIs A and B in Fig. 4. The display window sizes are [0, 1.5] and [0, 2].

To further evaluate the advantages of the proposed method, the difference material images with respect to the ground truth are shown in Fig. 6, and the display window width in terms of CT numbers of lung and head were [-2640, 639] and [-3459, 1459] HU, respectively. We observed that the difference images generated by the direct matrix inversion and iterative decomposition had significant errors, whereas the other three methods had fewer errors. Regarding the difference images of soft tissue generated by the FCN method, many blotchy artifacts appeared and are likely to be treated as image features, which might disturb tissue identification in some cases. Butterfly-Net can suppress this type of artifact to some extent, and our proposed DIWGAN can obtain the highest-quality images of soft tissue. In terms of the difference images of the bone, the results demonstrate that our proposed method can improve the accuracy significantly.

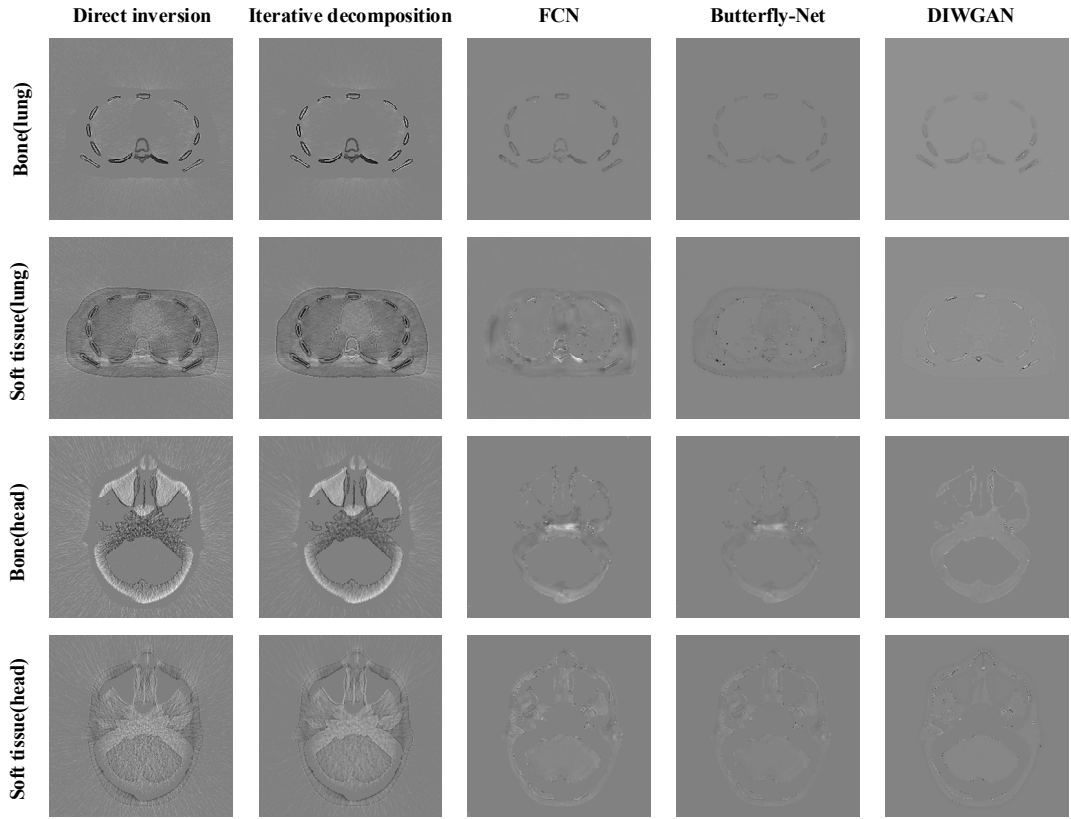

Fig. 6. Difference images between ground truths and decomposed material images. The display window sizes of lung and head are [-1, 1] and [-1.5, 1.5], respectively.

To quantitatively evaluate the material decomposition accuracy of all the methods, the RMSE, PSNR, and SSIM were calculated over the entire decomposed images. The evaluation results are listed in Table 2. Owing to the simple structure of the bone, the RMSE and PSNR obtained by the FCN, Butterfly-Net, and our proposed method were similar to those obtained using direct matrix inversion and iterative decomposition. In addition, the SSIM, an indicator used frequently to measure the similarity between two images, proved that the bone images generated by the proposed DIWGAN were the most similar to the ideal images among all the competitors. In terms of the soft-tissue images, some noise and artifacts still remained in the direct matrix inversion and iterative decomposition results, which resulted in inaccurate measured values. The FCN method can remove most of the noise and artifacts; however, it cannot easily discriminate noise or image features in edge areas. This can be confirmed from the larger RMSE compared with that of the proposed method. Compared with Butterfly-Net, the RMSE of the soft-tissue images generated by the DIWGAN decreased by 0.01 g/ml. The PSNR and SSIM of the soft-tissue images reached 31.43 dB and 0.9987, respectively, demonstrating the better decomposition accuracy and image quality of the proposed method. Additionally, the average RMSEs, PSNRs, and SSIMs of both lung and head data in all of the test sets are shown in Fig. 7. Cases 1 and 2 represent slices of bone and soft tissue from the lung, respectively, whereas Cases 3 and 4 represent those from the head, respectively. The results further proved the superior robustness of the proposed DIWGAN over the other decomposition methods.

**Table 2**

Quantitative evaluation results of different decomposition methods.

| Material | Index | Direct inversion | Iterative decomposition | FCN | Butterfly-Net | DIWGAN |
|---|---|---|---|---|---|---|
| Bone(lung) | RMSE(g/cm³) | 0.19 | 0.18 | 0.07 | 0.07 | **0.06** |
| | PSNR(dB) | 24.51 | 25.47 | 34.00 | 34.81 | **35.29** |
| | SSIM | 0.9939 | 0.9938 | **0.9963** | 0.9962 | **0.9963** |
| Soft tissue (lung) | RMSE(g/cm³) | 0.16 | 0.16 | 0.08 | 0.07 | **0.06** |
| | PSNR(dB) | 22.47 | 23.71 | 27.47 | 28.62 | **30.12** |
| | SSIM | 0.9896 | 0.9899 | 0.9959 | 0.9961 | **0.9968** |
| Bone(head) | RMSE(g/cm³) | 0.26 | 0.21 | 0.13 | **0.11** | **0.11** |
| | PSNR(dB) | 22.65 | 22.74 | 31.25 | 32.26 | **34.52** |
| | SSIM | 0.9880 | 0.9884 | 0.9990 | **0.9991** | **0.9991** |
| Soft tissue (head) | RMSE(g/cm³) | 0.26 | 0.20 | 0.11 | 0.10 | **0.08** |
| | PSNR(dB) | 21.47 | 21.65 | 30.80 | 31.21 | **31.43** |
| | SSIM | 0.9871 | 0.9879 | **0.9987** | **0.9987** | **0.9987** |

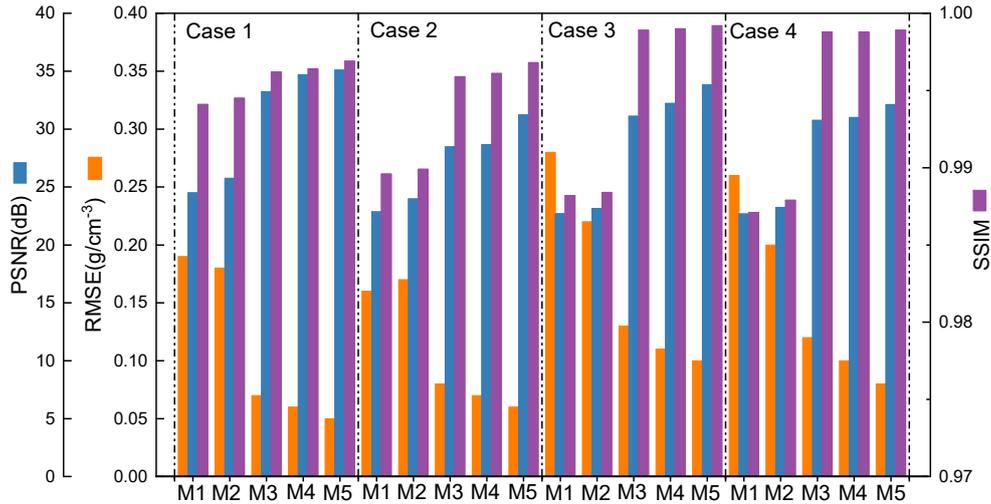

Fig. 7. Average indicators of test datasets by different methods. Note that M1, M2, M3, M4, and M5 represent the direct inversion, iterative decomposition, FCN, Butterfly-Net, and DIWGAN methods, respectively.

To quantitatively demonstrate the statistical advantages of the DIWGAN, the mean and SD of the ROIs marked with dotted red rectangles in Fig. 4 were calculated, and the results are shown in Table 3. Compared with direct matrix inversion, the DIWGAN reduced the noise SD of the selected ROIs in the decomposed images by 69.1%, 86.2%, and 79.5%. Meanwhile, compared with the iterative decomposition method, the noise SD reduced by 65.2%, 85.6%, and 60.0%. As for the results generated by the FCN, our proposed method exhibited better performances, i.e., noise SD reduction by 33.0%, 61.7%, and 35.8% for three ROIs. In terms of Butterfly-Net, the proposed method reduced the noise SD by 22.3%, 42.5%, and 21.2%. In addition, the profiles of a fine structure in the decomposed soft-tissue image under the red line in Fig. 8 (a) are shown in Fig. 8 (b) to demonstrate the distribution of the linear attenuation coefficient in each voxel from different methods. The DIWGAN was the most similar to the ground truth among all the competitors, further proving the advantages of the proposed method.

**Table 3**

A list of mean and SD on the different ROIs generated by different methods.

| Methods | ROI 1 | ROI 2 | ROI 3 |
|---|---|---|---|
| Direct inversion | 1.3344±0.0236 | 2.5716±0.1354 | 2.6024±0.0254 |
| Iterative decomposition | 1.3345±0.0210 | 2.5825±0.1299 | 2.4908±0.0130 |
| FCN | 1.3313±0.0109 | 2.8369±0.0488 | 2.1961±0.0081 |
| Butterfly-Net | 1.3292±0.0094 | 2.8853±0.0325 | 2.1987±0.0066 |
| DIWGAN | **1.3254±0.0073** | **2.9694±0.0187** | **2.1789±0.0052** |

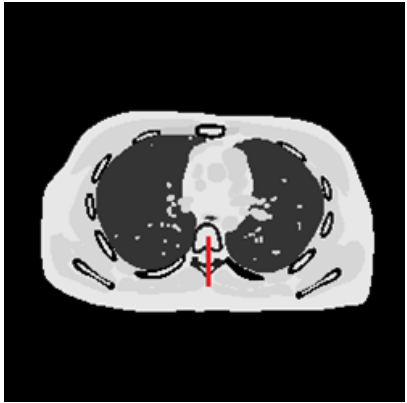

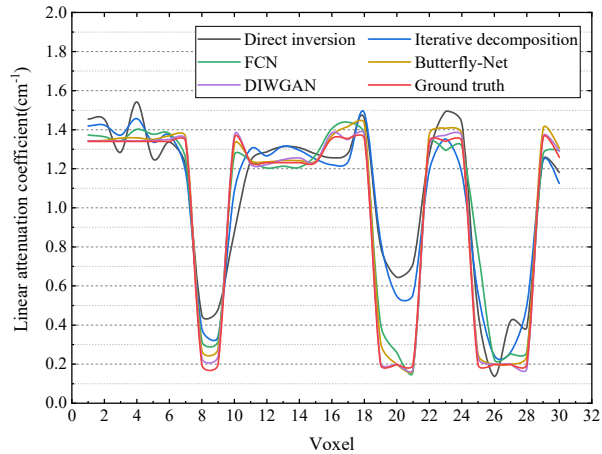

(a)              (b)

Fig. 8. Profiles of decomposition results under red line via different decomposition methods.

## 3.C. Real data test

To evaluate the effectiveness of our proposed method in real scanned data, mouse data were used to test the trained material decomposition model. To adapt the spectral range and scan protocols of the mouse datasets, the training data used in the experiment were regenerated with a pixel size of 110 μm × 110 μm under 36–52 and 52–80 keV energy bins. In addition, because the reconstructed mouse images contained a significant amount of noise, we increased the noise level of the regenerated training data, and the same amount of data was employed to retrain the decomposition model. After all the loss functions converged to a minimum, the reconstructed images of the mouse, as shown in Fig. 9 (a1, a2), were used to test the network performance. Figures 9 (b1–f1) and (b2–f2) show the decomposition results of the bone and soft tissue from different methods, respectively. We observed that the decomposition performances of the direct and iterative decompositions were poor. Noise was amplified in the decomposed soft-tissue images, and a large amount of tissue information was lost. In the learning-based methods, noise was suppressed significantly, and most of the tissue structures were preserved. As shown in Figs. 9 (d2, e2, and f2), the bone image generated by the DIWGAN was the most similar to the original structure in the reconstructed images. However, some detailed tissue structures were lost, particularly in the decomposed soft-tissue images; in our opinion, this was caused by the inconsistent noise distribution between the simulation and real data. In addition, the blooming artifact of the bone remained in the soft-tissue image generated by learning-based methods, which was discovered through the bright aliasing around the high-contrast boney structure in Figs. 9(d1, e1, and f1). Nevertheless, the results still demonstrated the decomposition potential of the DIWGAN in real scanned data, and the decomposition accuracy can be further improved if the model is trained with real scanned data.

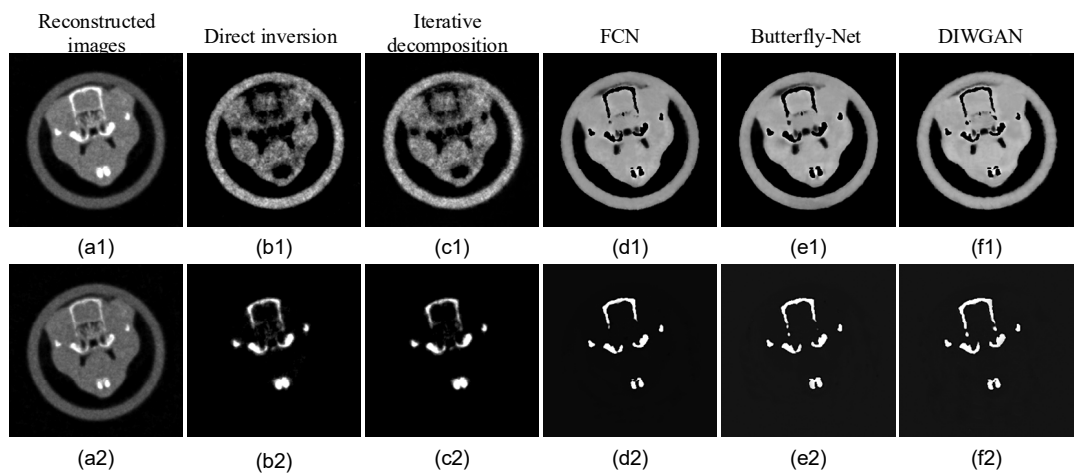

Fig. 9. Decomposition results of mouse. (a1) and (a2) represent reconstructed images from low- and
high-energy bins with display window size [0, 0.5] cm$^{-1}$. (b1–f1) and (b2–f2) are decomposed soft
tissue and bone images generated by different methods with display window size [0, 3.5], respectively.

### 3.D. Ablation experiment of hybrid loss function

To further illustrate the advantages of the proposed hybrid loss function, the effects
of different loss function combinations were compared, and the results are shown in
Fig. 10. A fine structure with significant gradient changes was selected to demonstrate
the decomposition accuracy and edge preservation. The profile of the decomposed soft-
tissue image under the red line in Fig. 10(a) is shown in Fig. 10(b) to demonstrate the
distribution of the linear attenuation coefficient in each voxel. As shown, the material
decomposition performance was poor when only the $L_1$ loss was used. When the $L_1$ and
edge losses were used simultaneously, the linear attenuation coefficient in the image
edge voxels was more accurate, but it can be further improved compared with the
ground truth. Among all the comparators, the hybrid loss including $L_1$, $L_{edge}$ and $L_{adv}$
achieved the best performance with a value the most similar to the ground truth. This
may be because the $L_1$ and edge losses are both pixel-wise objective functions, and the
network can only learn the pixel-level correlation between the reconstructed and
material-specific images. The adversarial loss enables the modeling of explicit pixel-
wise objective functions to be avoided, and a rich similarity metric is learned to
distinguish real data from and fake ones; therefore, features beyond the pixel level in
images can be learned and hence more realistic results can be obtained.

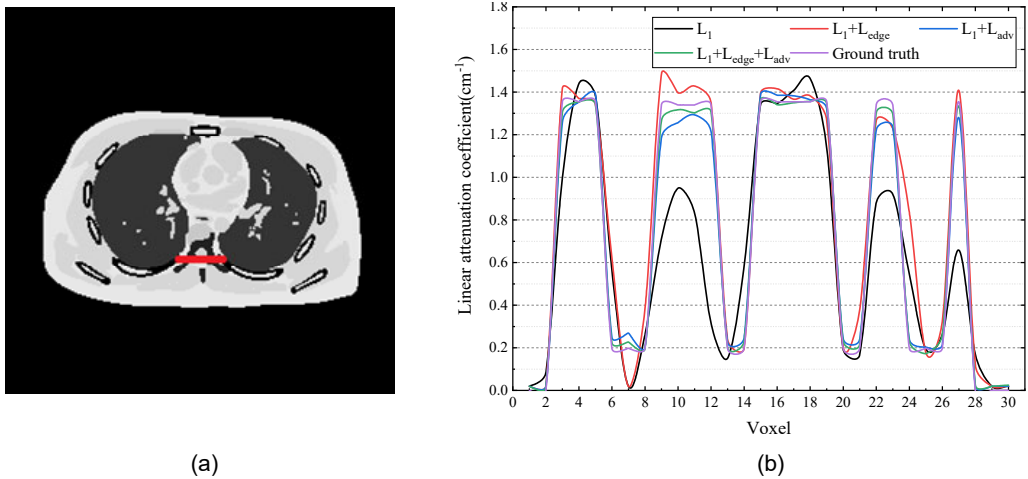

(a)        (b)

Fig. 10. Profiles of fine structure in soft-tissue image with different loss functions.

## 4. DISCUSSION:

The experimental results suggest that the material decomposition accuracy can be improved by the proposed data-driven approach. The high-quality images of bone and soft tissue generated by the DIWGAN demonstrate that noise and beam-hardening artifacts were successfully suppressed through adversarial training. The noise distribution and spectral correlation from different energy bins were learned using two interactive generators, and the feedback from the discriminators prevented smoothing in the material-specific images.

Practically and theoretically, training the network with only adversarial loss may cause missing diagnostic information. The hybrid loss used in the proposed method significantly affected the decomposition performance. The $L_1$ loss can ensure that the reconstructed images are mapped in a pixel-wise manner to the corresponding material images, and the edge enables causes the network to provide material-specific images with clear edges. In addition, when training the two generators synchronously, we discovered that they were always trapped in a local optimum, which degraded the decomposition performance of the trained model. Hence, a selector was employed to solve this problem, which was central to network training. By comparing different material decomposition methods, we discovered that the learning-based methods outperformed the traditional approaches. This might be because some noise has the same linear attenuation coefficient as the tissues. They were difficult to be distinguished by the direct matrix inversion and iterative decomposition methods, which leads to the noise magnification after decomposition. On the contrary, owing to the reference images, the noise and artifacts can be easily removed during model training. Based on the experimental results generated by the FCN, Butterfly-Net, and DIWGAN in this study, barely any noise and artifacts were detected, but the performances of the FCN and Butterfly-Net in the structure edges were unsatisfactory, which may be limited by their loss function. Hence, instead of minimizing only one mean-based loss, a hybrid loss along with adversarial training demonstrated significant potential in improving the network ability in terms of edge preservation. The experimental results confirmed that combining different losses according to the learning tasks is effective for further improving the performances of learning-based methods.

However, some issues exist in the proposed method. First, the parameters of the selector, which were used to balance the training of two generators, were selected via numerous experiments, and they were sensitive to datasets of different types and sizes. A possible solution is to design a network for the selector to learn from the distance between the ground truth and the generated images of two generators, and then define one that should be trained in the next iteration. Furthermore, the training data used in our experiment were obtained from only one type of dual-energy CT detector, which could not guarantee satisfactory performances for other spectral CT detectors. Hence, extended datasets from different types of detectors are required to further improve the robustness of this method.

## 5. CONCLUSIONS:

An image-based material decomposition method, known as the DIWGAN, was described and evaluated in this study. Two interactive generators were used to generate the corresponding material-specific images, and both spectral and spatial correlations were exploited by sharing feature maps. In addition, the hybrid loss enabled the generators to yield noise-free images with clear structure information, and the selector ensured that the losses of the generators converged to a minimum. The results demonstrated the effectiveness of the proposed DIWGAN, which could be beneficial for image-domain material decomposition. For our future study, we will adapt this method to different spectral CT detectors by using the corresponding data to train the decomposition model. In addition, three generators and the corresponding discriminators can be used in the network architecture such that it can be applied to three-material decompositions. Each generator is responsible for synthesizing one specific material and the feather exchanging can also be realized in the same way used in our proposed DIWGAN.

## ACKNOWLEDGEMENTS

This work was funded by the National Natural Science Foundation of China through the grant 61674115. Authors would like to thank Prof. Segars from Duke University for providing the phantom datasets, and we are also grateful to Prof. Anthony Butler and Hannah Prebble from MARS Bioimaging Ltd for sharing the mouse datasets.

## CONFLICT OF INTEREST STATEMENT

The authors have no relevant conflicts of interest to disclose.

## REFERENCES:

1.  Rubin GD. Computed tomography: Revolutionizing the practice of medicine for 40 years. *Radiology*. 2014;273(2):S45-S74. doi:10.1148/radiol.14141356

2.  Eiber M, Holzapfel K, Frimberger M, et al. Targeted dual-energy single-source CT for characterisation of urinary calculi: Experimental and clinical experience. *Eur Radiol*. 2012;22(1):251-258. doi:10.1007/s00330-011-2231-2

3.  Sukovic P, Clinthorne NH. Penalized weighted least-squares image reconstruction for dual energy X-ray transmission tomography. *IEEE Trans Med Imaging*. 2000;19(11):1075-1081. doi:10.1109/42.896783

4.  Niu T, Dong X, Petrongolo M, Zhu L. Iterative image-domain decomposition for dual-energy CT. *Med Phys*. 2014;41(4):041901. doi:10.1118/1.4866386

5.  Harms J, Wang T, Petrongolo M, Niu T, Zhu L. Noise suppression for dual-energy CT via penalized weighted least-square optimization with similarity-based

regularization. *Med Phys*. 2016;43(5):2676-2686. doi:10.1118/1.4947485

6. Xu Q, Yu HY, Mou XQ, Zhang L, Hsieh J, Wang G. Low-dose X-ray CT reconstruction via dictionary learning. *IEEE Trans Med Imaging*. 2012;31(9):1682-1697. doi:10.1109/TMI.2012.2195669

7. Chandarana H, Megibow AJ, Cohen BA, et al. Iodine quantification with dual-energy CT: Phantom study and preliminary experience with renal masses. *Am J Roentgenol*. 2011;196(6):W693-W700. doi:10.2214/AJR.10.5541

8. Saito H, Noda K, Ogasawara K, et al. Reduced iodinated contrast media for abdominal imaging by dual-layer spectral detector computed tomography for patients with kidney disease. *Radiol Case Reports*. 2018;13(2):437-443. doi:10.1016/j.radcr.2018.01.028

9. Primak AN, Fletcher JG, Vrtiska TJ, et al. Noninvasive Differentiation of Uric Acid versus Non-Uric Acid Kidney Stones Using Dual-Energy CT. *Acad Radiol*. 2007;14(12):1441-1447. doi:10.1016/j.acra.2007.09.016

10. Alvarez RE, MacOvski A. Energy-selective reconstructions in X-ray computerised tomography. *Phys Med Biol*. 1976;21(5):733-744. doi:10.1088/0031-9155/21/5/002

11. Foygel Barber R, Sidky EY, Gilat Schmidt T, Pan X. An algorithm for constrained one-step inversion of spectral CT data. *Phys Med Biol*. 2016;61(10):3784-3818. doi:10.1088/0031-9155/61/10/3784

12. Lehmann LA, Alvarez RE, Macovski A, et al. Generalized image combinations in

dual KVP digital radiography. *Med Phys*. 1981;8(5):659-667. doi:10.1118/1.595025

13. Xue H, Zhang L, Chen Z, Li L. A correction method for dual energy liquid CT image reconstruction with metallic containers. *J Xray Sci Technol*. 2012;20(3):301-316. doi:10.3233/XST-2012-0339

14. Li L, Li R, Zhang S, Zhao T, Chen Z. A dynamic material discrimination algorithm for dual MV energy X-ray digital radiography. *Appl Radiat Isot*. 2016;114:188-195. doi:10.1016/j.apradiso.2016.05.018

15. Wu W, Yu H, Chen P, et al. DLIMD: Dictionary learning based image-domain material decomposition for spectral CT. arXiv:1905.02567, 2019.

16. Xu Q, Xing L, Xiong G, Elmore K, Min J. SU-E-I-41: Dictionary Learning Based Quantitative Reconstruction for Low-Dose Dual-Energy CT (DECT). *Med Phys*. 2015;42(6Part6):3250-3251. doi:10.1118/1.4924038

17. Li M, Zhao Y, Zhang P. Accurate iterative FBP reconstruction method for material decomposition of dual energy CT. *IEEE Trans Med Imaging*. 2019;38(3):802-812. doi:10.1109/TMI.2018.2872885

18. Li Z, Ravishankar S, Long Y, Fessler JA. DECT-MULTRA: Dual-Energy CT Image Decomposition With Learned Mixed Material Models and Efficient Clustering. *IEEE Trans Med Imaging*. 2019;39(4). doi:10.1109/tmi.2019.2946177

19. Wang G. A perspective on deep imaging. *IEEE Access*. 2016;4:8914-8924. doi:10.1109/ACCESS.2016.2624938

20. Chen H, Zhang Y, Zhang W, et al. Low-dose CT via convolutional neural network. *Biomed Opt Express*. 2017;8(2):679. doi:10.1364/boe.8.000679

21. Zhang Z, Liang X, Dong X, Xie Y, Cao G. A Sparse-View CT Reconstruction Method Based on Combination of DenseNet and Deconvolution. *IEEE Trans Med Imaging*. 2018;37(6):1407-1417. doi:10.1109/TMI.2018.2823338

22. Xie S, Zheng X, Chen Y, et al. Artifact Removal using Improved GoogLeNet for Sparse-view CT Reconstruction. *Sci Rep*. 2018;8(1):1-9. doi:10.1038/s41598-018-25153-w

23. Kallenberg M, Petersen K, Nielsen M, et al. Unsupervised Deep Learning Applied to Breast Density Segmentation and Mammographic Risk Scoring. *IEEE Trans Med Imaging*. 2016;35(5):1322-1331. doi:10.1109/TMI.2016.2532122

24. Badea CT, Holbrook M, Clark DP. Multi-energy CT decomposition using convolutional neural networks. *Medical Imaging 2018: Physics of Medical Imaging. International Society for Optics and Photonics*.2018;105731O. doi:10.1117/12.2293728

25. Xu Y, Yan B, Zhang J, Chen J, Zeng L, Wang L. Image Decomposition Algorithm for Dual-Energy Computed Tomography via Fully Convolutional Network. *Comput Math Methods Med*. 2018;2018. doi:10.1155/2018/2527516

26. Zhang W, Zhang H, Wang L, et al. Image domain dual material decomposition for dual-energy CT using butterfly network. *Med Phys*. 2019;46(5):2037-2051. doi:10.1002/mp.13489

27. Goodfellow IJ, Pouget-Abadie J, Mirza M, et al. Generative adversarial nets. *Adv. Neural Inf. Process. Syst.* 2014: 2672–2680.

28. Wolterink JM, Leiner T, Viergever MA, Išgum I. Generative adversarial networks for noise reduction in low-dose CT. *IEEE Trans Med Imaging*. 2017;36(12):2536-2545. doi:10.1109/TMI.2017.2708987

29. Mao X, Li Q, Xie H, Lau RYK, Wang Z, Smolley SP. Least Squares Generative Adversarial Networks. *Proceedings of the IEEE International Conference on Computer Vision*. 2017:2813-2821. doi:10.1109/ICCV.2017.304

30. Nowozin S, Cseke B, Tomioka R. f-GAN: Training Generative Neural Samplers using Variational Divergence Minimization. *Adv Neural Inf Process Syst*. 2016:271-279.

31. Arjovsky M, Chintala S, Bottou L. Wasserstein generative adversarial networks. *34th International Conference on Machine Learning, ICML 2017*. 2017:298-321.

32. Gulrajani I, Ahmed F, Arjovsky M, Dumoulin V, Courville A. Improved Training of Wasserstein GANs. *Adv Neural Inf Process Syst*. 2017:5768-5778.

33. Liu MY, Breuel T, Kautz J. Unsupervised image-to-image translation networks. *Adv Neural Inf Process Syst*. 2017:701-709.

34. Zhang Z, Li M, Yu J. D2PGGAN: Two Discriminators Used in Progressive Growing of GANS. *IEEE International Conference on Acoustics, Speech and Signal Processing - Proceedings*. 2019:3177-3181.

35. Shi Z, Li J, Li H, Hu Q, Cao Q. A Virtual Monochromatic Imaging Method for

Spectral CT Based on Wasserstein Generative Adversarial Network With a Hybrid Loss. *IEEE Access*. 2019;7:110992-111011. doi:10.1109/access.2019.2934508

36. Yang G, Yu S, Dong H, et al. DAGAN: Deep De-Aliasing Generative Adversarial Networks for Fast Compressed Sensing MRI Reconstruction. *IEEE Trans Med Imaging*. 2018;37(6):1310-1321. doi:10.1109/TMI.2017.2785879

37. Szczykutowicz TP, Chen GH. Dual energy CT using slow kVp switching acquisition and prior image constrained compressed sensing. *Phys Med Biol*. 2010;55(21):6411-6429. doi:10.1088/0031-9155/55/21/005

38. Ronneberger O, Fischer P, Brox T. U-net: Convolutional networks for biomedical image segmentation. *Lecture Notes in Computer Science (Including Subseries Lecture Notes in Artificial Intelligence and Lecture Notes in Bioinformatics)*. 2015:234-241. doi:10.1007/978-3-319-24574-4_28

39. Zareapoor M, Zhou H, Yang J. Perceptual image quality using dual generative adversarial network. *Neural Computing and Applications*, 2019: 1-11.

40. Nair V, Hinton GE. Rectified linear units improve Restricted Boltzmann machines. *ICML 2010 - Proceedings, 27th International Conference on Machine Learning*. 2010:807-814.

41. You C, Yang Q, Shan H, et al. Structurally-sensitive multi-scale deep neural network for low-dose CT denoising. *IEEE Access*. 2018;6:41839-41855. doi:10.1109/ACCESS.2018.2858196

42. Segars WP, Sturgeon G, Mendonca S, Grimes J, Tsui BMW. 4D XCAT phantom

for multimodality imaging research. *Med Phys*. 2010;37(9):4902-4915. doi:10.1118/1.3480985

43. Shi Z, Yang H, Cong W, Wang G. An edge-on charge-transfer design for energy-resolved x-ray detection. *Phys Med Biol*. 2016;61(11):4183-4200. doi:10.1088/0031-9155/61/11/4183

44. Siddon RL. Fast calculation of the exact radiological path for a three dimensional CT array. *Med Phys*. 1985;12(2):252-255. doi:10.1118/1.595715

45. Walsh MF, Nik SJ, Procz S, et al. Spectral CT data acquisition with Medipix3.1. *J Instrum*. 2013;8(10):P10012. doi:10.1088/1748-0221/8/10/P10012

46. Kingma DP, Ba JL. Adam: A method for stochastic optimization. arXiv: 1412.6980,2015.

47. Abadi M, Agarwal A, Barham P, et al. TensorFlow: Large-Scale Machine Learning on Heterogeneous Distributed Systems. *GPU Comput Gems Emerald Ed*. 2016: 277-291. doi:1603.04467